\RequirePackage{lineno}
\documentclass[english, preprint,preprintnumbers,amsmath,amssymb]{revtex4}

\usepackage{hyperref}
\usepackage{cleveref}

\usepackage{graphicx}
\usepackage{dcolumn}
\usepackage{bm}
\usepackage[english]{babel}
\usepackage{epstopdf}
\usepackage{epsfig,color,amsmath}

\usepackage{gensymb}

\usepackage{nomencl} 
\makenomenclature

\newcommand{\bea}{\begin{eqnarray}}
\newcommand{\eea}{\end{eqnarray}}

\begin{document}

\title{ Mechanical Interaction Between Cells Facilitates Molecular Transport  }
\author{David Gomez$^{1}$, Sari Natan$^{1}$, Yair Shokef$^{1,2*}$ and Ayelet Lesman$^{1*}$}
\affiliation {$^{1}$School of Mechanical Engineering, Tel Aviv University, Tel Aviv, Israel}
\affiliation {$^{2}$Sackler Center for Computational Molecular and Materials Science, Tel Aviv University, Tel Aviv, Israel}
\affiliation {*ayeletlesman@tauex.tau.ac.il}
\affiliation {*shokef@tau.ac.il}






\begin{abstract}


\textit{In vivo}, eukaryotic cells are embedded in a matrix environment, where they grow and develop. Generally, this extracellular matrix (ECM) is an anisotropic fibrous structure, through which macromolecules and biochemical signaling molecules at the nanometer scale diffuse. The ECM is continuously remodeled by cells, via mechanical interactions, which lead to a potential link between biomechanical and biochemical cell-cell interactions. Here, we study how cell-induced forces applied on the ECM impacts the biochemical transport of molecules between distant cells. We experimentally observe that cells remodel the ECM by increasing fiber alignment and density of the matrix between them over time. Using random walk simulations on a 3D lattice, we implement elongated fixed obstacles that mimic the fibrous ECM structure. We measure both diffusion of a tracer molecule and the mean first-passage time a molecule secreted from one cell takes to reach another cell. Our model predicts that cell-induced remodeling can lead to a dramatic speedup in the transport of molecules between cells. Fiber alignment and densification cause reduction of the transport dimensionality from a 3D to a much more rapid 1D process. Thus, we suggest a novel mechanism of mechano-biochemical feedback in the regulation of long-range cell-cell communication.




\end{abstract}

\maketitle

\section{Introduction}
\label{sec:Introduction}

The extracellular matrix (ECM) is a complex fibrous microenvironment that provides the scaffold for cells and tissues. Cells continuously interact and modify the ECM structure by degrading and reassembling the ECM structural blocks \cite{Davis,Iozza,Kielty,Kim,Page-McCaw,Frantz}. Additionally, cells interact mechanically with the ECM by applying active forces by their internal actin-myosin machinery \cite{Kim,Legate,Trubelja,Mak}. Such cell-induced forces deform and reorganize the ECM fibers, and specifically cause the matrix to align and densify between neighboring cells \cite{Jansen,Korff,Shi,Kim2,Stopak,Notbohm,Notbohm1,Liu}. Consequently, through matrix-mediated mechanical interactions and mechanosensing mechanisms, distant cells communicate and regulate key cellular processes such as orientation \cite{Winer}, migration \cite{Korff,Wang1}, and cancer invasion \cite{Shi,Han,Spilla}. Moreover, cells integrate information of the ECM's architecture and mechanical properties via interactions between fiber elements and transmembrane proteins \cite{Kim,Frantz}. For example, structural changes in the ECM environment are translated by the transmembrane protein integrin, into a series of biochemical processes inside the cell that influence cell spreading, motility, and differentiation \cite{Kim2,Discher,Janmey}.


The \textit{in vivo} ECM is a milieu of proteins, polysaccharides, and growth factors that diffuse and transfer biochemical signals between cells \cite{Kim,Clause,Bokel,Fassler}. Regulation of the transport of such molecules and how they locate their targets directly influences cell behavior \cite{Winer,Korff,Shi,Clause,Eliceiri,Ross}. The molecular dynamics at the ECM microenvironment has been at the heart of extensive theoretical and experimental studies \cite{Kihara,Clague,Jiao,Leddy,Tourell,Stylianopoulos,Ramanujan,Netti}. {\color{black}Many have examined the role of fiber orientation on the diffusion coefficient of macromolecules (e.g., dextran molecule), in tissues and hydrogels, together with Monte Carlo simulations models \cite{Leddy,Tourell,Stylianopoulos}.} Aligned environments support higher diffusion parallel to the preferred orientation of fibers, as compared to isotropic fiber distributions {\color{black}(without any preferred orientation) \cite{Stylianopoulos,Leddy,Erikson,Tourell,Clague,Jiao}}. The diffusion coefficient parallel to the fiber orientation remains effectively invariant in the case fibers are aligned, whereas its perpendicular component decreases due to the multiple encounters of the diffusing molecule with the fibers. Similarly to the effects described in the context of molecular crowding \cite{Zimmerman2,Morelli,Gomez}, the influence of the elongated fiber elements on molecular diffusion depends on the intrinsic properties of the molecules, such as their size and charge \cite{Leddy,Tourell,Stylianopoulos}. Despite considerable experimental and theoretical efforts, scientists still lack a detailed understanding of the possible effects of ECM fiber anisotropy on the effectiveness of biochemical transport processes. Most of the studies on the cell-ECM interactions have been done either on ECM remodeling by both mechanical and chemical processes, or on ECM molecular interactions. {\color{black}In this study, we experimentally show that neighboring cells mechanically remodel the fibrous matrix between them to form aligned and dense bands of fibers. Based on this experimental motivation, we propose a novel mechanism for cell-cell long-range interaction: we couple the ECM dynamic structure, with the induced effect on molecular dynamics in anisotropic environments. We demonstrate by a computational model that increased alignment and densification of fibers results in strong speed-up of molecular transport throughout a possible mechanical-biochemical feedback process.}

We employ a lattice model to address the effects of fiber orientation and densification on the transport of molecules between neighboring cells. In our model, molecular motion takes place in tortuous channels of {\color{black}connected voids} that emerge as the fiber occupation fraction increases. We analyze how fiber density affects channel continuity and study its effects on molecular diffusion between a pair of cells. As we approach the percolation threshold, {\color{black}the fiber density at which connectivity of empty voids gets disrupted,} we observe a drastic change in the molecular dynamics, due to the critical changes in the topology of the system. {\color{black}Also, we explore how fiber orientation and densification affect target finding (i.e., a neighboring cell). In our model, the first-passage time (FPT) is the time that it takes for a molecule to find its target for the first time. We thus run simulations to obtain the mean first-passage time (MFPT) of target finding.} We observe that as fiber alignment and volume occupation fraction increase, the MFPT sharply decreases due to a confinement effect that supports a faster molecular transport. We study how such effects depend on intrinsic molecular features such as the molecule-to-fiber size ratio. Our experimental observation of cells embedded in biological gels suggests that the ECM between cells reaches values in alignment and densification, at which our model predicts a substantial effect on molecular transport. As such, we propose a novel mechanism of cell-cell communication, in which cell-induced forces increase fiber alignment and densification, which in turn lead to the emergence of channel-like fiber configurations that support a faster transport of molecules. Our lattice model aims to create a clear conceptual understanding of the effects of ECM structural remodeling on diffusion and on the biochemical communication between pairs of cells, rather than providing an extensive computational description of the ECM remodeling and molecular-ECM interactions. Our model can, however, be extended to include additional types of fiber-molecule interactions as well as other fiber geometries.

\section{Methods}

\subsection{Experiments}

\subsubsection{Cell preparation and microscopy }

Cells culture: Swiss 3T3 fibroblasts stably transfected with GFP-actin (obtained as gifts from S. Fraser, University of Southern California, Los Angeles, CA) were cultured in DMEM medium supplemented with 10$\%$ fetal bovine serum, nonessential amino acids, sodium pyruvate, L-glutamine and 1$\%$ PSN (100 units/ml Penicillin, 100 $\mu$g/ml Streptomycin, and 12.5 units/ml) in a 37$^\circ$C humid incubator. 

\subsubsection{3D fibrin gel preparation }

Actin-GFP 3T3 fibroblast cells (8000 cells) were mixed with 10 $\mu$l of a 20 U/ml thrombin solution (Omrix Biopharmaceuticals). 10 $\mu$l of 10 mg/ml fluorescently labeled fibrinogen (Omrix Biopharmaceuticals) was prepared. The fibrinogen suspension was placed on a No. 1.5 coverslip in a 35-mm dish (MatTek Corporation), and mixed gently with the thrombin-cells suspension. The resulting fibrin gel was placed in the incubator for 20 min to fully polymerize, after which warm medium was added to cover the gel.

\subsubsection{Fibrin gel labeling }

Alexa Fluor 546 carboxylic acid, succinimidyl ester (Invitrogen) was mixed with fibrinogen solution in a 7.5:1 molar ratio for 1 hour at room temperature and then filtered through a HiTrap desalting column (GE Healthcare) packed with Sephadex G-25 resin to separate the unreacted dye. The labeled fibrinogen was then mixed with thrombin and cells to create labeled, cell-loaded fibrin gels.

\subsubsection{Time-lapse microscopy }

Pairs of cells were imaged by confocal microscopy, Zeiss 880 using a 40X water immersion lens (Zeiss) equipped with 30mW Argon Laser (wavelengths 488, 514 nm).  Throughout imaging, cells were maintained in a 37$^\circ$C/5$\%$ CO$_2$ incubation chamber allowing for normal cell growth. Confocal z-stacks capturing fibroblasts cells were acquired every 15min.

\subsubsection{Image analysis}

Confocal images of the fibrous matrix between cell pairs were analyzed using ImageJ (NIH, Bethesda, MD; http://imagej.nih.gov/ij). The distribution of fiber orientation was calculated by the ImageJ plug-in, OrientationJ \cite{Puespoeki}. We analyzed the 2D images overtime for 120 minutes, after visually detecting cell-cell interaction by band formation, see figure 1. We calculated the density over time and obtained the orientation distribution and the nematic order parameter in the region between the cell pairs and in a control region, far from the cells, see figure 1. The general definition of the nematic order parameter $S$ is given by:  $S=  \frac{d \langle \cos^{2} \theta \rangle -1}{d-1} $, with $d$ the dimensionality of the system and $\theta$ the angle between the fiber orientation and the preferred direction of orientation \cite{Mercurieva}. $S=0$ accounts for an isotropic environment, whereas $S=1$ accounts for a completely aligned system. Experimentally, we analyze 2D slices, $d=2$, and the preferred direction of orientation is defined by the axis between the centers of the cells. Thus, $S= \langle \cos{2 \theta} \rangle $ \cite{Mercurieva,Alvarado}.

\subsection{Computational model}

We study the transport of molecules in ECM environments using kinetic Monte Carlo simulations of particles moving on a 3D simple cubic lattices with periodic boundary conditions {\color{black} in all three principal cubic directions}. The simulation box has a volume of $V=L_x\times L_y \times L_z$ lattice sites. The system contains three types of molecules: a tracer molecule (representing, for example, a growth factor secreted by one cell), a target molecule (in this case a growth factor receptor on the membrane of a neighboring cell) and fibers. We set the molecule, and the target each to occupy a single lattice site. The target is taken to be static and is positioned at $x=(L_x/2)$, $z=(L_z/2) $ and $y=(3L_y/4) $. We set the initial position of the molecule to be at $x=(L_x/2)$, $z=(L_z/2) $ and $y=(L_y/4)$, as shown in figure 2. Hence, when considering a lattice of dimension $L_x=L_y=L_z=100$, the volume is $V=10^{6}$ lattice sites and the molecule-target initial distance is $\lambda=50$ lattice sites along the $y$-axis. {\color{black} In our simulations, the cell-cell distance $\lambda$ is normalized with respect to the molecule size, which we set to be equal to 1. Thus, for the case of $\lambda=50$, and molecule size of 0.01 $\mu$m, the cell-cell distance equals 0.5 $\mu$m. }


ECM fibers are implemented as linear arrangements of lattice sites running through the whole lattice length and oriented along one of the three principal lattice axes. {\color{black}Fibers occupy a volume fraction $\phi =1 -  (1 - Nw^{2}p_{x}/L_yL_z ) (1 - Nw^{2} p_{y}/L_xL_z ) (1 - N w^{2} p_{z}/L_xL_y )$, with $N$ the total number of fibers and $w$ the fiber thickness, see figure 2B. Note that as $w$ increases, a smaller number of fibers is needed to get the same occupation fraction. Fiber thickness $w$ is also normalized to the molecule size. Thus, a fiber with a thickness of $w$ has a cross-section larger by a factor of $w^2$ than the molecule, as shown for the cases $w=$ 2 and 4 in figure 2B.} In our implementation, each fiber is set to run along the $x,y$ or $z$ axes with probability $p_x,p_y$ and $p_z$, respectively. In the simulations, we characterize different fiber configurations by substituting $d=3$ into the aforementioned general expression for the nematic order parameter: $S=\frac{3 \langle \cos^{2} \theta \rangle -1}{2} $ \cite{Mercurieva,Stephen}. In this case, $\theta$ is defined as the angle between the fiber orientation and the $y$-axis, which we set to be the preferred direction of alignment. Thus, when $p_x=p_y=p_z=1/3$, the system is isotropic and the order parameter has a value of $S=0$. In the fully aligned scenario, when $p_x=p_z = 0$ and $p_y=1$, the order parameter equals $S=1$, as shown in figure 2. {\color{black} Here, the direction of fiber alignment is the same as band formation, i.e. along the cell-cell center of mass.} We also considered systems in which diagonal fibers ($\theta=45^\circ$) are included. We observed no substantial change in the results (data not shown). {\color{black}Using our model, we examined the effect of four different parameters, including fiber density, alignment, and thickness, and cell-cell distance. We modified these parameters to be within the physiological range shown in Table 1.}



\begin{table}[ht]
\centering
\begin{tabular}{| c | c | c | }
\hline

 & \textit{In vitro}  & Our model  \\  [-1.8ex]
 &(hydrogels) &\\ 
\hline
Density & 0.02-0.7 Fibrin   \cite{Leonidakis}  & 0-0.9  \\  [-1.8ex]
(volume fraction)  & 0.002-0.1 Collagen \cite{Kreger,Kuntz,Reese}  & \\ 
Fiber thickness &  0.01-0.3 $\mu$m (Collagen-Fibrin) \cite{Stevens,Wade}  & 0.01-1 $\mu$m \\  
Cell-Cell distance & 14-80 $\mu$m Collagen  \cite{Kim2,Liu2}  & 0.1-200 $\mu$m   \\  
Alignment  & 0.18-0.7  \cite{Vader,Lee,Riching,Taufalele} & 0-1 \\ [-1.8ex]
(NOP)  &  & \\ 
\hline
\end{tabular}
\caption{{\color{black}Relevant biological parameters compared to the parameters used in the computational model. In the model, cell-cell distance is in the range of $10 \leq \lambda \leq 2000$, and fiber thickness $1 \leq w \leq 10$ (both normalized values in respect to the molecule size of 1). Therefore, a typical GF size, approximately 0.01 $\mu$m corresponds to fibers 0.01-0.1  $\mu$m thick and cell-cell distances of 0.1-20  $\mu$m. For an extracellular vesicle, about 0.1 $\mu$m in diameter, fiber thickness ranges 0.1-1 $\mu$m and the cell-cell distance 1-200 $\mu$m. NOP is the nematic order parameter ($S$).}}
\label{table:nonlin}
\end{table}




{\color{black}At each simulation time step, which defines the unit of time, the molecule advances to one of its six neighboring sites with a probability of $1/6$, provided the desired site is empty. If a fiber occupies the desired site, the move is rejected and the time increases by a unit of time.} Thus, steric effects are the only considered interactions, and on an empty lattice ($\phi=0$), the molecule diffuses with a diffusion coefficient $D_{\alpha}=D_{0}/3$ in each cubic direction, being $\alpha=x,y,z$ and $D_0= 1/6$ the total 3D diffusion coefficient at $\phi = 0$. As fiber occupation fraction increases, diffusion is hindered, thus $D(\phi)<D_0$. For our lattice with periodic boundary conditions, we obtain $D(\phi)$ by waiting until the mean-square displacement grows linearly with time, i.e., the dynamics turns diffusive. {\color{black}The latter is obtained by running simulations over $10^6$ time steps}. For the case with aligned fibers ($S=1$), for which the fibers are directed along the y-axis, we also run similar simulations on a 2D square lattice of size $L_x\times L_z$ with periodic boundary conditions. {\color{black}For the MFPT calculations, we acquire the time that it takes for the molecule to find its target, and then we start a new realization of the simulation. The statistics are performed for $10^{4}$ finding events. Here, if the molecule finds its target, a molecule-target complex is immediately formed. Thus, the reaction is taken to be diffusion-limited.}


\section{Results}

\subsection{Experimental characterization of fiber alignment and densification}

Cells can exert forces and deform their surroundings \cite{Wang,Guidry,Sawhney,Shokef,Xu,Spilla,Malandrino}. In 3D fibrous gels, such deformations can induce densified regions between distant cells \cite{Korff,Kim2,Stopak,Notbohm,Shi,Ma,Sopher}. We show here the induction of directed matrix remolding created between neighboring cells in a 3D setting. Fibroblast cells expressing GFP-Actin were seeded in a fibrous biological gel (fibrin). Within a few hours from seeding, the cells deform the matrix, occasionally forming a thick band of fibers between them, see figure 1A. In figure 1B, histograms of the angle of fiber orientations at different times during band formation are presented. As time increases, more of the fibers reorient around zero, thus demonstrating the alignment of the fibers between the cells. Far from the cells (control), the angle distribution remains flat, see figure 1C. Moreover, we calculate the nematic order parameter of the fibers in two different sample locations. In the band area between the cells, the order parameter increases as a function of time to a high degree of alignment $S \approx 0.7$ (figure 1D). In the control area, far from the intercellular band, no alignment is observed as seen in figure 1D. Also, we quantified the intensity in the area between the cells, demonstrating that it gradually increases up to two-fold relatively to the control in about two hours (figure 1E). Since the gel is uniformly labeled, the increase in intensity directly infers on the rise in fiber occupation fraction in the band region. Our observations of matrix densification and alignment between cells are highly heterogeneous and dynamic: not every pair of cells leads to the formation of a band, and the time of band initiation differs between pairs of cells. Additionally, it is worthy to note that image acquisition was made only after recognizing ECM remodeling by cells. Therefore, at time zero in figure 1, D and E, substantial alignment and densification have already taken place. {\color{black}Our experiments demonstrate that cells induce gradual, time-dependent increase in fiber densification and alignment in the matrix between them.} The latter motivates to study the effect of fiber orientation and densification has on molecular diffusion. {\color{black}We hypothesize that the aligned and dense fibers (`bands') generated between distant cells may improve the transport of molecules traveling in this remodeled area (i.e., molecules that are secreted by the interacting cells). We next analyze through a computational model, the effects of fiber volume occupation fractions and orientations on the dynamics of a diffusing molecule, therefore examining the link between mechanical and biochemical cellular interaction in ECM.}



\subsection{Diffusion in Simulated Fibrous Environments}

To quantify the effects of ECM remodeling on the transport of molecules, we generate lattices with different fiber occupation fractions $\phi$, distributed in the two extreme nematic order parameter values, $S = 0$ (isotropic) and $S=1$ (fully aligned along the $y$-axis), {\color{black}and in an intermediate value $S=0.5$}, as shown in figure 2. We then measure their effect on the molecule's diffusion, see figure 3. We plot in figure 3A the diffusion coefficient as a function of occupation fraction of fibers for the two values of isotropy. In the aligned system ($S=1$), the diffusion coefficient decays roughly linearly in occupation fractions $\phi < 0.4$. At $\phi  \approx 0.4$, $D(\phi)/D_0$ reaches a value of $1/3$ and remains constant for larger values of $\phi$. This is because the system reaches percolation at $\phi_C^{2D} \approx 0.4$ \cite{Stauffer}. In the aligned system, fibers run only along the $y$-axis, thus changes in the fiber occupation fraction influence the motion only in the $x-z$ cross-section, as schematically shown in figure 3B. Therefore, the aligned fiber case can be mapped to a 2D site percolation problem on the square lattice \cite{Stauffer}. We thus study, by running 2D lattice simulations, the evolution of the mean-square displacement $(\langle r^{2} \rangle)$ as a function of $\phi$ and plot in figure 3C the ratio $\langle r^{2} \rangle / t$ as a function of time. In this representation, normal diffusion yields a line of slope zero with diffusion coefficient $D_{\perp}(\phi)$ given by the limiting value of $\langle r^{2} \rangle / t$, for large $t$ \cite{Sokolov,Saxton1,Saxton2,Weiss,Saxton3}, as shown by the black dashed lines in figure 3C. As $\phi$ increases toward $\phi_C^{2D}$, tortuous channels emerge in the x-z plane and hinder diffusion. At short times, $\langle r^{2} \rangle$ evolves as $t^{\alpha}$, with $\alpha < 1$ the subdiffusion exponent. At long times, correlations in diffusion disappear, and the molecule recovers its diffusive behavior. Note that as $\phi$ increases, and approaches $\phi_C^{2D}$, the subdiffusion-diffusion crossover time becomes larger until it eventually diverges at $\phi_C^{2D}$. For $\phi>\phi_C^{2D}$, the diffusive behavior across the $x-z$ plane vanishes, causing the perpendicular component of the diffusion coefficient $D_{\perp}(\phi) = (D_x(\phi)+D_z(\phi))/2$ to be zero. Along the direction of fiber alignment, the parallel diffusion coefficient $D_{\parallel}(\phi) = D_y(\phi)=1/3$ remains constant, since diffusion in the $y$ direction is not affected by the addition of fibers, as shown in figure 3D. Consequently, for $ \phi \geq \phi_C^{2D}$, in the x-z plane, the molecule is confined, and its diffusion is effectively reduced in dimensionality from a 3D process to a 1D random walk along the preferred direction of fiber alignment. 




Now, fibers isotropically distributed are equally positioned in all three lattice directions ($x,y$, and $z$), i.e., $S=0$. As the fiber occupation fraction increases, complex, labyrinth-like structures emerge and hinder the diffusion. At a critical fiber occupation fraction of $\phi_C^{3D}$, the continuity of the empty space in the system is completely lost, and the molecule becomes trapped. This exact model has been numerically studied in the context of linear holes being removed from the lattice and has been characterized with the drilling percolation density $\phi_C^{3D} \approx 0.75$ \cite{Kantor,Schrenk}. This 3D percolation problem is nontrivial due to the complicated correlations between the fiber blocks in the linear arrangement of the fibers, and therefore, its effects on diffusion have not, to our knowledge, been studied. As expected, as the fiber occupation fraction increases, the diffusion coefficient decays roughly linearly until reaching zero at $\phi_C^{3D} \approx 0.75$ (figure 3A). For fiber occupation fractions $\phi \geq 0.75$, the diffusion coefficient remains zero due to the complete 3D caging of the molecule. In this fiber configuration, steric effects equally hinder diffusion in all directions of motion due to the isotropic distribution of fibers. Therefore, the perpendicular and the parallel components of the diffusion coefficient are both affected as the fiber volume fraction increases, see figure 3D. {\color{black}For intermediate fiber orientation values $S=0.5$, the diffusion coefficient decreases to zero at a fiber density, which is larger than for the fully isotropic case $\phi=0.9$.}



The ECM is a microenvironment composed of molecules with different sizes and with fibers that vary in thickness, typically between 10-300 nm \cite{Stevens,Wade}. Therefore, a natural parameter to modulate is the fiber thickness of $w$. We run simulations for different values of $w$ and plot in figure 4A the effects of fiber volume fraction on the diffusion coefficient for the case with aligned fibers ($S=1$). {\color{black}Note that as fiber thickness increases, a smaller number of fibers is needed to obtain the same volume fraction $\phi$.} In general, thicker fibers have a weaker impact on molecular diffusion than the case where the molecule and the fibers have the same thickness. Essentially, correlations in the positions of the building blocks of the thick fibers generate larger holes between the fibers, enabling the molecule to escape. Consequently, molecular diffusion is unhampered in the holes created by the thicker fibers. A similar effect has been observed in the reduction of depletion forces that big crowders have over small tracer molecules in the context of molecular crowding \cite{Zimmerman2,Morelli,Gomez}. For the aligned configuration, our 3D implementation allows us to determine the critical fiber volume fraction $\phi_C^{2D}$ for each fiber thickness, at which the system percolates in the $x-z$ plane. This is determined by obtaining the value of $\phi$ at which the diffusion coefficient reaches a value of $D(\phi)=1/3$. As fiber thickness increases, the percolation fiber volume fraction $\phi_C^{2D}$ increases, as shown in the inset of figure 4A.

To further analyze our system, we determine the critical fiber volume fraction $\phi_C^{3D}$ at which the diffusion coefficient $D(\phi)$ decays to zero in the isotropic configuration $S=0$, for different fiber thicknesses $w$. $\phi_C^{3D}$ can be equivalently computed by obtaining the probability that the molecule finds its target, i.e., $P_{F}$. This event can only take place when the molecule and the target belong to the same cluster of continuously-connected empty lattice sites between which the molecule can hop. If the cluster of vacant lattice sites does not contain both the target and the source locations, it implies that the molecule will never be able to find its target. Also, the molecule becomes trapped and $D(\phi)=0$. We ran structural calculations with periodic boundary conditions along the $x$ and $y$ axes for different values of $\phi$ and $w$. We obtained $P_F$ by calculating the probability for the molecule to reach a lattice site in the top of the lattice ($z=L$), given that its initial position is at the bottom of the simulated box ($z=0$). For our finite system size, we obtain $\phi_C^{3D}$ by plotting $P_{F}$ as a function of $\phi$, for different system sizes $L$, as shown in figure 4B. {\color{black}$\phi_C^{3D}$ is defined as the fiber volume fraction, at which $P_{F}$ for different system sizes cross each other. Using this definition of the fiber density threshold, $P_F$ exhibits a sharper transition as the system size ($L$) increases \cite{Newman,Newman1}}. For fiber volume fractions  $\phi \geq\phi_C^{3D}$, {\color{black}the voids continuity} from the bottom to the top of the lattice is entirely disrupted. For fibers with thickness $w=1$, we recover the obtained critical volume fraction $\phi_C^{3D} \approx 0.75$. In the inset of figure 4B, we plot $\phi_C^{3D}$ for different values of $w$. As fibers become thicker, the critical fiber volume fraction shifts towards higher values; hence, more of the thicker fibers are required to cage the molecule completely. {\color{black}We note that for each fiber thickness $w$, a similar plot of $P_{F}$ as a function of $\phi$ is obtained for three different system sizes, and $\phi_C^{3D}$ is defined. }





\subsection{Kinetics of Molecular Transport in Fibrous Environments}

Next, we explore the kinetics of the transport of the molecule towards its target. A quantity of key interest is the MFPT as a function of fiber thickness, fiber volume fraction, and network isotropy. The MFPT allows us to capture in detail the implications of emerging local fiber structures that can drive molecular transport far from diluted conditions. We start by considering three different degrees of isotropy, $S=0, 0.5, 1$, and modulate the fiber volume fraction $\phi$ for different fiber thicknesses, as presented in figure 5A. In case no fibers are present in the system, the average time the molecule needs to reach its target is given by MFPT$^{3D}=(kc)^{-1}$, where $k=\gamma D_0 a$ is the corresponding diffusion-limited finding rate \cite{Smoluchowski}, $\gamma$ a numerical prefactor which equals 4 in our cubic lattice,  $D_0=1/6$ the diffusion coefficient in the absence of fibers, $a$ the target size (taken here to be unity), and $c=1/V$ the molecule's concentration (since we consider a single molecule). Consequently, our results are system-size dependent. For $L_x = L_y = L_z = 100$, the MFPT for $\phi=0$ is MFPT$^{3D}= 3V/2=1.5 \cdot 10^{6}$, as shown in figure 5A (upper dash-dot line).



When fibers are all aligned ($S=1$), the average finding time follows, for all considered fiber thicknesses ($w$), a monotonic decrease as the volume fraction of fibers increases. In the case of $w=1$ and for fiber volume fractions $\phi < \phi_C^{2D} =0.4$, the MFPT decreases moderately due to the smaller available volume, see figure 5A. Here, the molecule's dynamics is diffusive and the MFPT scales with the system's volume, as can be observed in the trajectory for $\phi=0.3$ presented in figure 5B. Just above the percolation threshold at fiber volume fraction $\phi_C^{2D}$, a sharp decrease in the MFPT takes place due to reduction of dimensionality: the large number of aligned fibers generate channel-like structures that prevent the long-time diffusive behavior along the $x-z$ plane, allowing diffusion only along the preferred fiber orientation (y), as shown by the trajectory in figure 5B ($\phi=0.5$). As $\phi$ keeps increasing, the characteristic size of the channels becomes smaller, as shown in figure 5B ($\phi=0.6$), and the MFPT further decreases asymptotically approaching the value of MFPT$^{1D}$, denoted by the lower dash-dot line in figure 5A. For our system size, MFPT$^{1D}$ $=  \lambda^{2}/2D_0=3\lambda^{2} = 7500$, where $ \lambda^{2} $ accounts for the squared initial molecule-target distance, which in our simulations is equal to 50 lattice sites.


Since the MFPT is system-size dependent, and the distance between cells can vary significantly in biological systems, we plot in figure 5C the MFPT as a function of the cell-cell distance for aligned fibers. Here, we use $L_x=L_z=20$, modulate the cell-cell distance $\lambda=L_y/2$, and maintain periodic boundary conditions in all three directions. For short cell-cell distances and low $\phi$, the MFPT is effectively given by MFPT$^{3D}$ (red line) as it scales with the system's volume, thus linearly with $\lambda$, see figure 5C. As the cell-cell distance increases, the system becomes more elongated along the $y$-axis and MFPT$^{1D}$ (blue line) becomes more dominant, and it scales quadratically with $\lambda$. At a cell-cell distance of $\lambda^{*}=L_x \times L_z= 400$, MFPT$^{1D}$ and MFPT$^{3D}$ cross each other and beyond that distance MFPT$^{1D}$ dominates. Importantly, for cell-cell distances shorter than $\lambda^*=400$, reduction of dimensionality has a strong effect on the MFPT. For several cell-cell distances, we run simulations with $\phi=0, 0.3, 0.5, 0.6,$ and $0.8$. As $\phi$ increases, the MFPT becomes smaller and shifts towards the more efficient MFPT$^{1D}$, as indicated by the orange arrow. The effect of reduction of dimensionality decreases as the cell-cell distance increases until vanishing at large cell-cell distances, as shown by the purple data.

In the isotropic case with $S=0$, the finding time follows a non-monotonic behavior for all fiber sizes as the fiber volume fraction increases. In the case $w=1$, the finding time exhibits a mild decrease until $\phi \approx 0.3$. {\color{black}For these values of $\phi$, the finding time slightly decreases due to the smaller available volume}. As $\phi$ keeps increasing, diffusion of the molecule becomes more hindered in all directions of motion, causing the molecule to spend longer times close to its initial position. As a consequence, the finding time starts to increase from the observed minimum at a fiber density $\phi \approx 0.3$. For higher fiber volume fractions ($\phi  > 0.4$), the {\color{black}continuity of voids} between the molecule's initial position and its target starts to vanish, and the diffusion coefficient further decreases. Thus, the MFPT becomes large. As the fiber volume fraction approaches the critical threshold $\phi_C^{3D} \approx 0.75$ (figure 4C), {\color{black}void continuity} between the molecule and its target disappears, and the MFPT of target finding starts to diverge. Varying fiber thickness does not change the qualitative description of the effects of fibers on the finding time, see figure 5A. Nevertheless, quantitative differences emerge due to the higher critical fiber volume fractions for the systems with thicker fibers. {\color{black}Note that for very thick fibers ($w=10$), a considerable decrease in the finding time is obtained at high fiber volume fractions. }

The non-monotonic behavior in the finding time is strengthened when considering an intermediate isotropy level, e.g., $S=0.5$ (gray data in figure 5A). Here, fiber positioning is not uniform, and more fibers are placed along the preferred direction ($y$-axis) than in the two perpendicular directions ($x$- and $z$-axes). Hence, similarly to the aligned case $(S=1)$, as fiber volume fraction increases (above approximately $\phi=0.55$), the positioning of fibers on the $x-z$ plane leads to the formation of a channel that confines diffusion along the $y$-axis. {\color{black}Nevertheless, in contrast to the aligned system, here, as the fiber volume fraction further increases ($\phi \geq 0.65$), fibers can perpendicularly run through the channel.} Thus, {\color{black}void channel continuity} starts to be disrupted, and the MFPT increases until the continuity is wholly lost, causing the finding time to diverge (figure 5A).




To further understand the complex behavior of the finding time, we study how the finding time varies for different values of isotropy at given fiber volume fractions. We thus create a color map shown in figure 6A of the finding time as a function of both $\phi$ and $S$ for fibers with thickness $w=1$. As expected, fiber alignment and fiber volume fractions $\phi \leq 0.4$, don't have a substantial impact on the finding time, see blue data in figure 6B. In isotropic configurations, large values of $\phi$ encage the molecule, leading to very long or diverging finding times, as shown in the top-left red area in figure 6A. The non-monotonic behavior of the finding time is observed at intermediate values of $\phi$ and $S$. Specifically, at isotropy values $0.4 \leq S \leq 0.55$, it is possible to follow the complicated form of the finding time as $\phi$ is modulated. In particular, for the case of $S=0.5$, figure 5A is recovered when modulating $\phi$.


For high enough fiber volume fractions, changes in isotropy have strong effects on the target finding time. In particular, the finding time can be tuned from long or divergent times (red region) to short finding times (blue zone), by increasing fiber alignment in the system at very high fiber volume fractions ($\phi \geq 0.7)$, see figure 6A. Note that the blue zone describes a 1D dynamics. At intermediate fiber volume fractions, e.g., $\phi = 0.5$ (green data in figure 6B), the finding time sharply decreases as fibers align along the preferred direction. When considering a higher fiber volume fraction of $\phi=0.6$, the qualitative behavior of the finding time remains the same. The molecule finds its target more quickly, as shown by the red data in figure 6B. A crucial result of our model is that the increment in the target finding efficiency is due to the reduction of dimensionality and the emergence of topological structures that support the transport of molecules between pairs of cells.

To directly measure the benefit of fiber alignment on molecular transport, we quantify the ratio of the MFPT in a perfect isotropic system MFPT$_{S=0}$, over the MFPT in a fully aligned environment MFPT$_{S=1}$. In figure 6C, we plot this ratio MFPT$_{S=0}$/MFPT$_{S=1}$ as a function of the fiber volume fraction for different fiber thicknesses. When the ratio is large, alignment is highly efficient, whereas if it approaches one, there is not a significant benefit induced by fiber alignment. As fibers become thicker, higher fiber volume fractions are needed to increase the efficiency of the transport of molecules. Interestingly, aligned fibers are more than 30 times more efficient than the randomly distributed fibers, even when fibers are ten times thicker than the molecule. Consequently, under physiological conditions, in which fibers are typically one order of magnitude thicker than the diffusing particles, alignment of fibers can have significant implications on the transport of molecules.

\section{Discussion}


Based on our experimental observations and theoretical modeling, we propose a novel signaling transduction mechanism, in which an interplay between biochemical reaction processes coupled to mechanical mechanisms, can lead to a more efficient long-range cell-cell communication, see figure 7. Although the understanding of biochemical signaling transduction and mechanical signaling processes have been at the heart of numerous theoretical and experimental studies \cite{Janmey,Sapir,Chen,Berg,Kaizu,Golkov,Ben-Yaakov}, most efforts have been devoted to understanding them separately. To our knowledge, our work is the first to address the interplay between mechanical and biochemical cell-cell communication.



We experimentally showed that cells could remodel the ECM environment by locally increasing fiber alignment and density between them. Our simulations predict that increments in fiber alignment and density in values that correspond to our experimental measurements can improve the transport of molecules in a manner that depends on the molecule-to-fiber size ratio.

Our computational modeling shows that as the fiber volume fraction and alignment increase, the MFPT for target finding drastically decreases (figure 6). Consistent with previous studies \cite{Stylianopoulos,Leddy,Tourell}, our results show that molecular diffusion is strongly affected in the perpendicular component of fiber orientation, but not along the parallel direction. Indeed, we showed that at high enough fiber volume fractions and alignment, typically above 0.5 and 0.6 respectively, the molecule's dynamics changes from a 3D to a 1D diffusion process resulting in very rapid transport of molecules. We postulate that this reduction of dimensionality potentially enhances molecular transport, and therefore, affects cell-cell biochemical communication.

In our experiments, we followed the dynamical transformation of our fibrous hydrogels in volume sections between (band) and far from (control) the cells. In particular, we experimentally observed that fibers between cells could become highly aligned, reaching nematic order parameter values of 0.5-0.7 in a couple of hours, as shown in figure 1. It is noteworthy that the experimental measurement of alignment was obtained by calculating the nematic order parameter ($S$) from 2D images. Thus, as shown in figure 5A, fibers perpendicularly crossing the band can hinder the transport of molecules, leading to the less efficient transport of molecules. Moreover, we estimate an increase in fiber densification between cells of more than two-fold based on the increase in fluorescence intensity. Before any effect of cell-induced remodeling, we expect our fibrous gels to have an initial volume fraction between $\phi=0.1$ and 0.5, as reported in a similar gel system \cite{Leonidakis}. We note that it is possible to tune the initial density of the gel by changing protein gel concentration to reach the desired effect on molecular transport. We thus expect ECM remodeling to positively modulate molecular transport efficiency as depicted by the black arrow in figure 6A.

In ECM environments, signaling molecules, like growth factors (GFs), regularly diffuse between distant cells targeting their receptors (growth factor receptor GFR) sitting at the cellular membrane of distant cells. Enhancement of GFs transport toward GFR  within the complex ECM structure can directly influence the internal biochemical signaling cascade, leading to different cellular responses, see figure 7. Specifically, when considering the insulin GF with a diameter of 6x$10^{-3}$ $\mu$m \cite{Parang}, and the ECM with fiber thickness around 0.1 $\mu$m \cite{Stevens,Wade}, we could expect a moderate increase in molecular transport. In our model, we work in dimensionless units by setting the molecule size to unity. Hence, fiber thickness in our simulations should be related to the experimental fiber thickness normalized by the molecule diameter. i.e., when considering the insulin GF,  $w=0.1$$\mu$m/6x$10^{-3}$$\mu$m $\approx 17$. {\color{black}In the current computational work, we considered excluded-volume interactions without taking into account the interaction that may exist between the molecules and fibers, for example, proteins that bind to the ECM fibers. Therefore, our model is more relevant for molecules which do not have an affinity to the ECM fibers. These can include, for example, the vascular endothelial GF (VEGF$_{120}$) that is freely diffusive in ECM microenvironments and links to the ECM weakly \cite{Martino,Ruhrberg,Park}. Also, the insulin-like GF has a lower affinity to the ECM fibers when present at high ionic salt concentration \cite{Jones}. Considering molecule-ECM binding interactions is expected to affect the model outcomes, and it is a natural extension of our model in future studies.} Furthermore, it was recently shown that cells release extracellular vesicles (EV) to the ECM environment \cite{Huleihel}. These EVs carry in their interior different molecules like cytokines, chemokines, and other proteins that play a role in cell signaling. When reaching neighboring cells, the EVs can be engulfed by the cells, internalizing then the EV content. The typical size of an EV is 30-100nm in diameter \cite{Huleihel,Nawaz}. Hence, the effects of ECM remodeling on the transport of EVs across the ECM are expected to be stronger and more determinant. {\color{black}We note that throughout this study, we coarse-grained the complex ECM geometry by the implementation of fibers on a discrete cubic lattice and the use of periodic boundary conditions. In real biological situations, fibers can be curved, highly disordered, and interconnected into networks \cite{Lindstroem}, and molecules can escape the analyzed volume between the cells. Future work will focus on more complicated fiber structures that better resemble physiological ECM networks, and on open boundary conditions.}




Given the widespread relation between mechanosensing and biochemical mechanisms, a possible mechano-biochemical feedback process could take place at the complicated focal adhesion interface, as shown in figure 7. Our results show that it is only through fiber densification and alignment that molecules can be more efficiently transported along the ECM to reach a nearby cell. {\color{black} Our model predicts a feedback mechanism between mechanical (cellular forces that remodel the fibers between cells) and biochemical (molecular transport) interactions. Possibly, the enhanced molecular transport and signaling can feedback to increase cell forces and ECM remodeling. For example,} GF-GFR interactions trigger downstream cellular signaling pathways at the focal adhesion, leading to integrin recruitment at the cell membrane and more mechanical interactions with the ECM fibrous structure \cite{Kim}. Also, our findings suggest that the transport of molecules in the ECM may have a broader significance for signaling transduction and other types of feedbacks may take place. For example, Lin \textit{et al.} showed that fibroblasts uptake to their interior proteins and other biomolecules by applying mechanical forces to the ECM, leading to different cellular responses, such a cellular migration \cite{Lin}. {\color{black}Thus, this study lays the foundation for the development of new experiments. For example, looking more closely on the transport of molecules within ECM remodeling between contractile cells. Or, exploring the possible feedback between the changes in mechanical properties of the remodeled ECM (stiffness) together with the transport of molecules, and its consequences on cell fate.}

\section{Acknowledgements}
We thank Gregory Bolshak, Rakesh Chatterjee, Yakov Kantor, Erdal C. Oguz, Ron Peled, Shlomi Reuveni and Raya Sorkin for helpful discussions. This work was partially funded by the Tel Aviv University postdoctoral program (D.G), by the US-Israel Binational Science Foundation (Y.S), by the Israel Science Foundation (Y.S. and A.L with grant numbers 968/16 and 1474/16, respectively), and the Israel Science Foundation-Israeli Centers for Research Excellence (A.L grant number 1902/12).

\printnomenclature[2.0cm]
\renewcommand\chaptername{Glossary}



\begin{center}
\begin{figure}
\centering
\includegraphics[width=17cm, height=10.5cm]{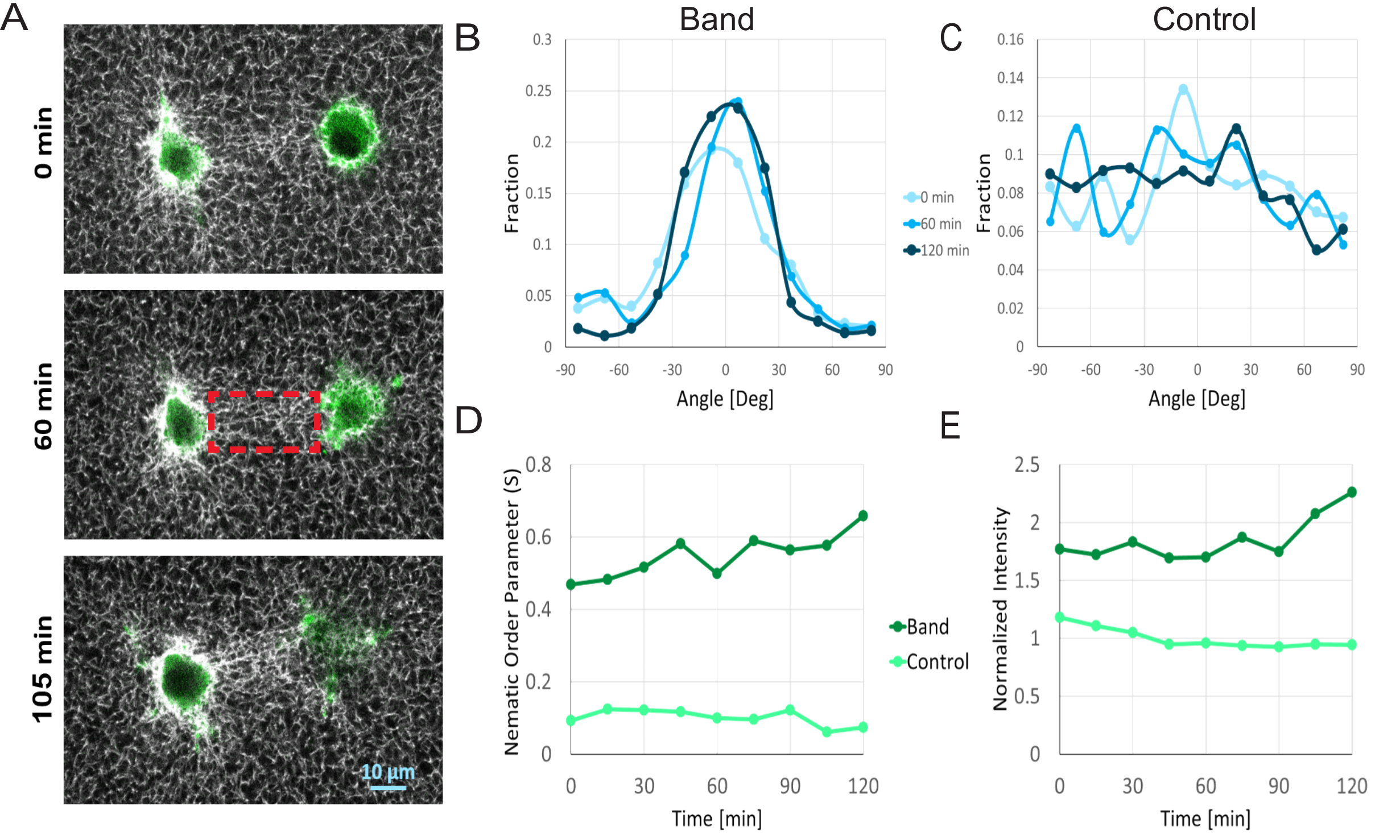} 
\caption{Band formation between two fibroblast cells labeled with GFP-Actin (green) embedded in a fibrous biological gel (gray). {\color{black}The band area between the cells was chosen to occupy the entire area between the cells (dashed red square), with a width that equals the cell diameter. The control area was chosen to be far from any nearby cells and had the same size as the band area}. We took high-resolution images only after identifying band formation (Time=0). Thus, at this initial time, fibers in the band already have some levels of alignment and densification.  A) Cells mechanically remodel the fibrous environment over time, by forming a band of dense matrix between them. B) In the region between the cells, fiber angle distribution peaks around zero as time increases. C) Fiber angle is uniformly distributed in regions far from the cells, showing the isotropic structure of the gel. D) The nematic order parameter ($S$) increases as a function of time in the area between the cells, whereas far from the cells $S$ remains close to zero. E)  Fiber intensity between cells shows a gradual increase as time evolves, indicating fiber densification.   }
\end{figure}
\end{center}

\begin{center}
\begin{figure}
\centering
\includegraphics[width=11cm, height=15cm]{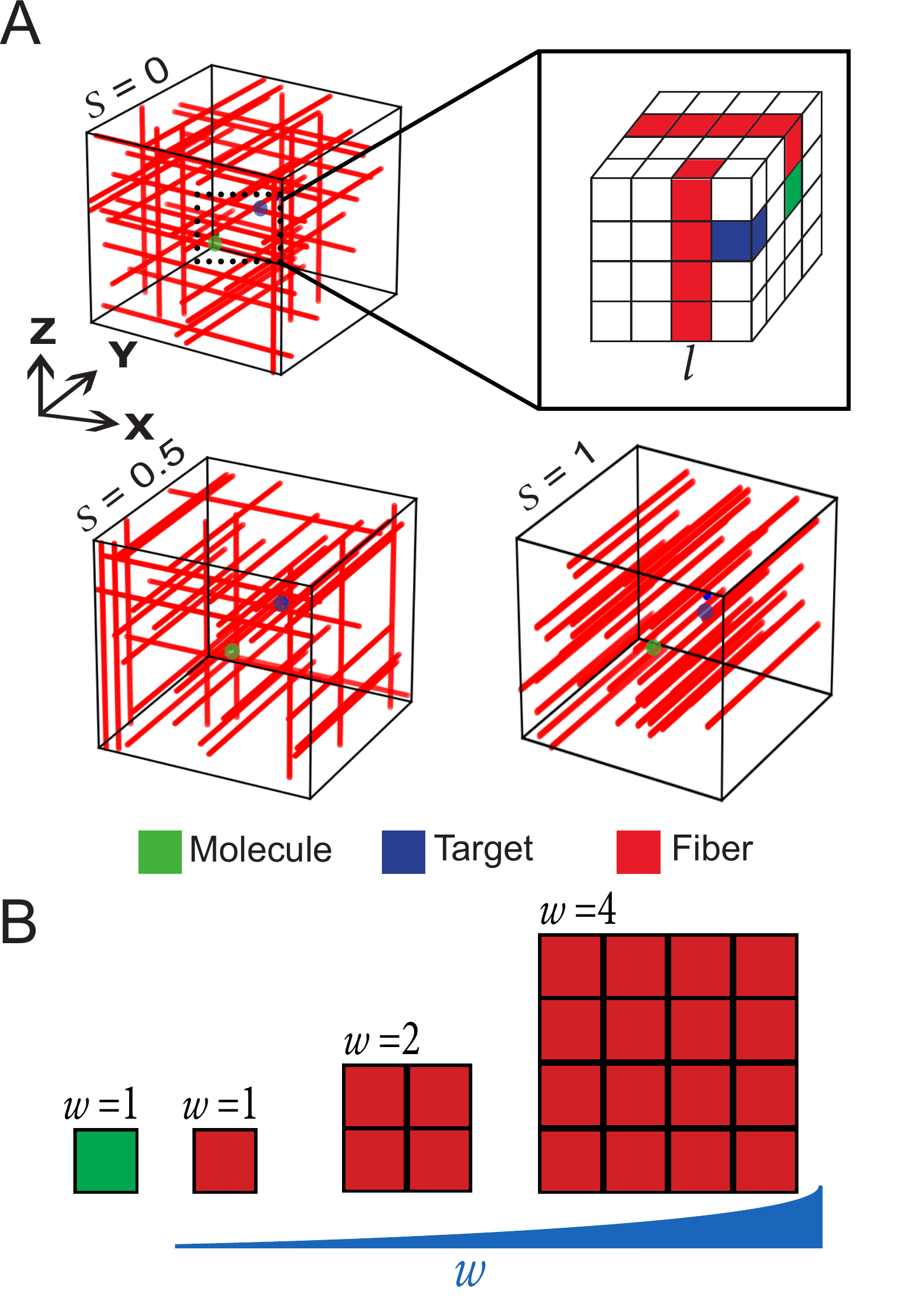} 
\caption{On-lattice model for molecular transport in fibrous environments. A molecule (green) diffuses by hopping to one of its six nearest-neighbor lattice sites, given the destination site is empty. Diffusion takes place until the molecule finds its target (blue). We implement the ECM environment by placing fiber elements (red) along with the three principal directions of the lattice, with proportions defining the alignment $S$ (see text). Here, the cases of isotropic fibers ($S=0$), partially aligned ($S=0.5$) and fully aligned ($S=1$) are shown. B) Schematic representation ($X$-$Z$ cross-section) of thicker fibers implemented in our lattice model.}

\end{figure}
\end{center}

\begin{center}
\begin{figure}
\centering
\includegraphics[width=12cm, height=12cm]{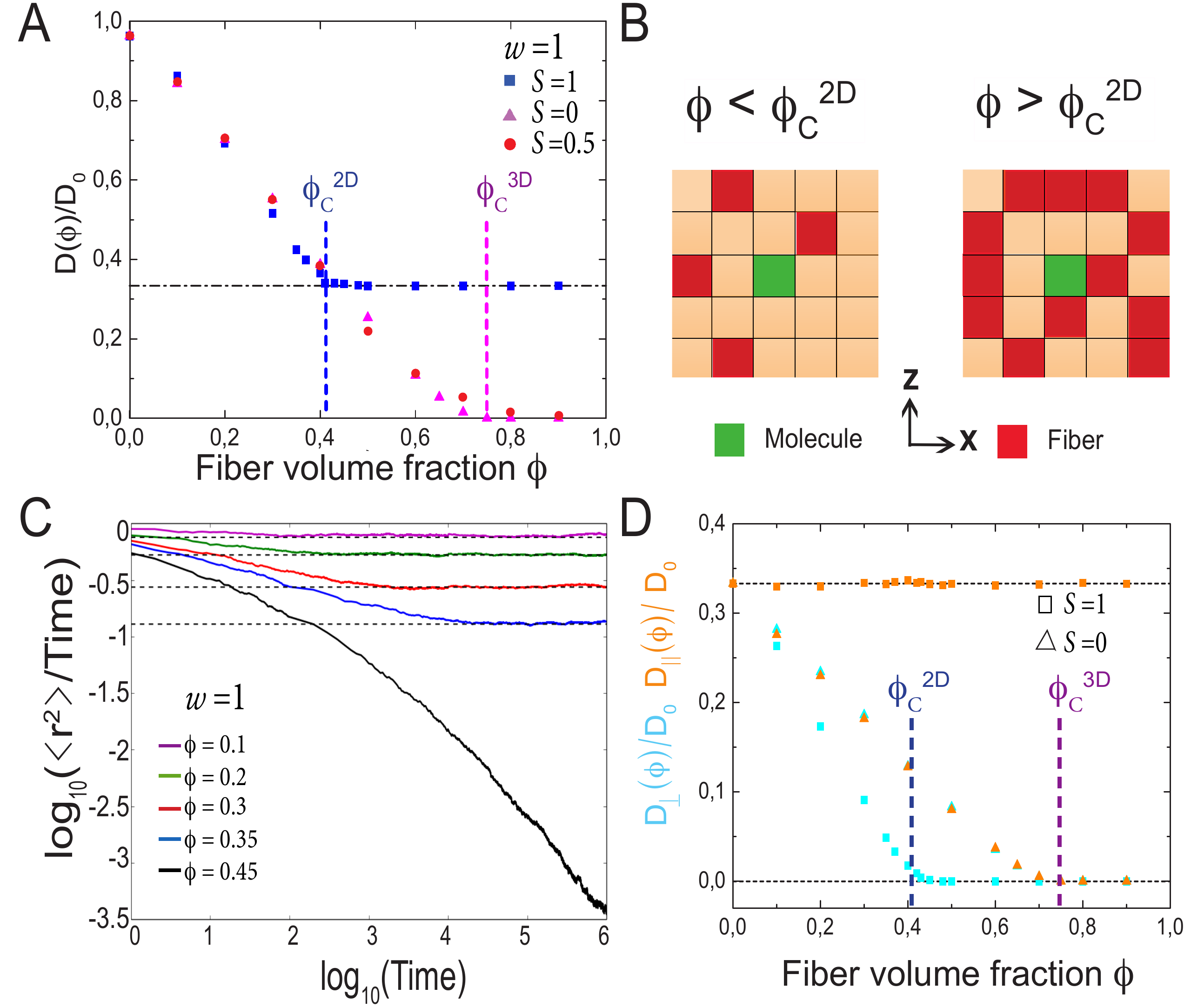} 
\caption{Effects of fibrous environments on molecular diffusion. A) $D(\phi)/D_0$ as a function of the fiber occupation fraction $\phi$ for the two extreme nematic order parameter values $S=0$ and $S=1$, and an intermediate value $S=0.5$. In the aligned system ($S=1$), the diffusion coefficient decays until reaching a constant value of $1/3$ at a fiber volume fraction $ \phi_C^{2D} \approx 0.4$. A system with fibers distributed with $S=1$ is equivalent to site percolation in a square lattice as shown in (B). When $S=0$, $D(\phi)/D_0$ decays until it reaches zero at a fiber volume fraction $ \phi_C^{3D} \approx 0.75$, corresponding to the critical volume fraction for drilling percolation. {\color{black}For $S=0.5$, $D(\phi)/D_0$ approaches zero at a higher fiber density $\phi \approx 0.9$.} C) 2D simulations of the temporal evolution of $\langle r^{2} \rangle$ for different values of $\phi$, for $S=1$. Horizontal lines correspond to diffusive behavior. D) $D_{\perp}(\phi) = (D_x(\phi)+D_z(\phi))/2)$ and $D_{\parallel}(\phi)= D_y(\phi)$, the perpendicular and parallel components of the diffusion coefficient as a function of $\phi$ for $S=0$ and $S=1$. All simulations are with system size $L_x = L_y = L_z = 100$. }
\end{figure}
\end{center}

\begin{center}
\begin{figure}
\centering
\includegraphics[width=14cm, height=7cm]{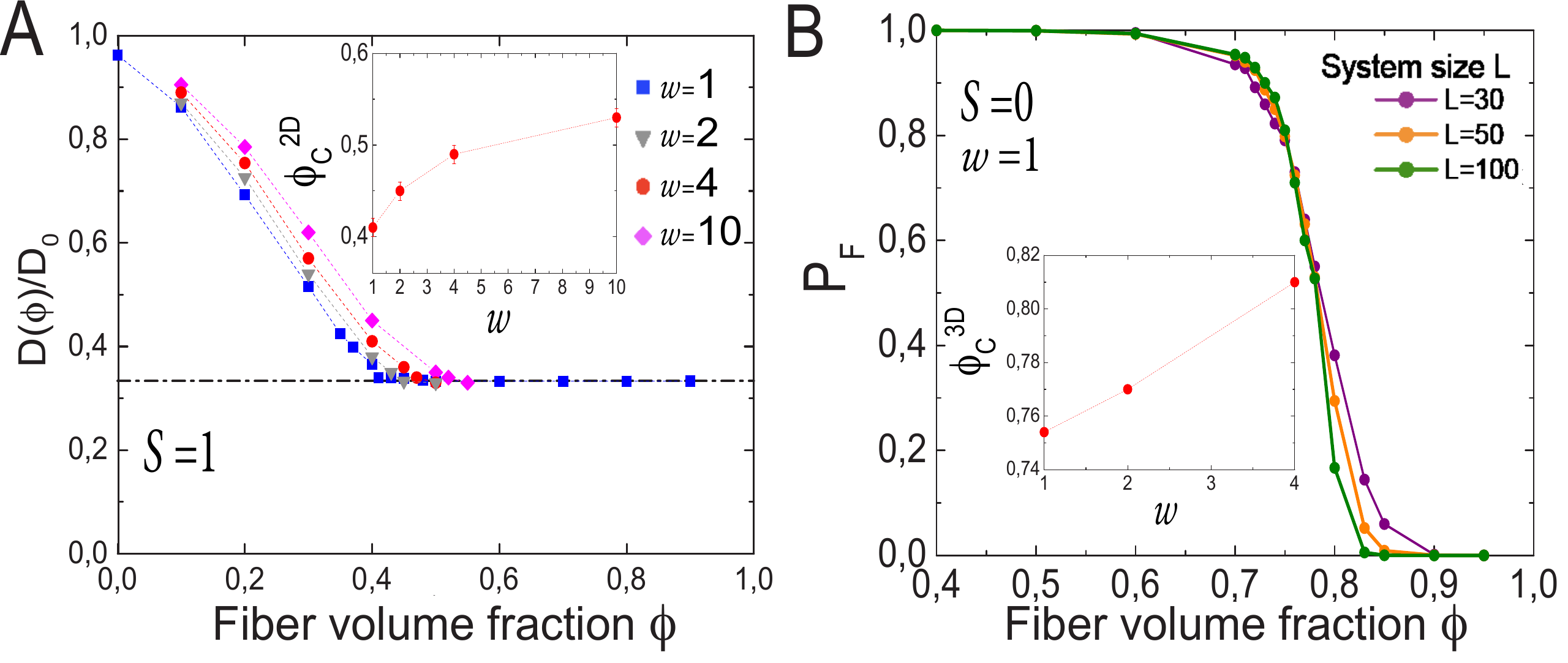} 
\caption{Impact of fiber thickness on molecular diffusion and percolation occupation fractions for the aligned and isotropic systems. The molecule size is kept constant and equal to 1. A) Diffusion coefficient as a function of $\phi$ for four different fiber thicknesses for the fully aligned system ($S=1$). As fiber thickness increases, molecular diffusion becomes less hindered, and the critical fiber volume fraction shifts towards higher values of $\phi$ (inset). Simulation parameters: $L_x = L_y = L_z = 100$. Error bars in inset account for the discrete increments in $\phi = 0.01$. {\color{black}B) P$_{F}$, the probability of reaching the top of the simulation box, provided a random walk starts from any position at the bottom, as a function of the fiber volume fraction $\phi$. Fibers are isotropically distributed ($S=0$). For each fiber thickness $w=1$, 2 and 4, $\phi_C^{3D}$ is defined as the density at which P$_{F}$ of the different system sizes cross each other. $\phi_C^{3D} $ becomes larger as $w$ increases, as shown in the inset. }}
\end{figure}
\end{center}

\begin{center}
\begin{figure}
\centering
\includegraphics[width=12cm, height=12cm]{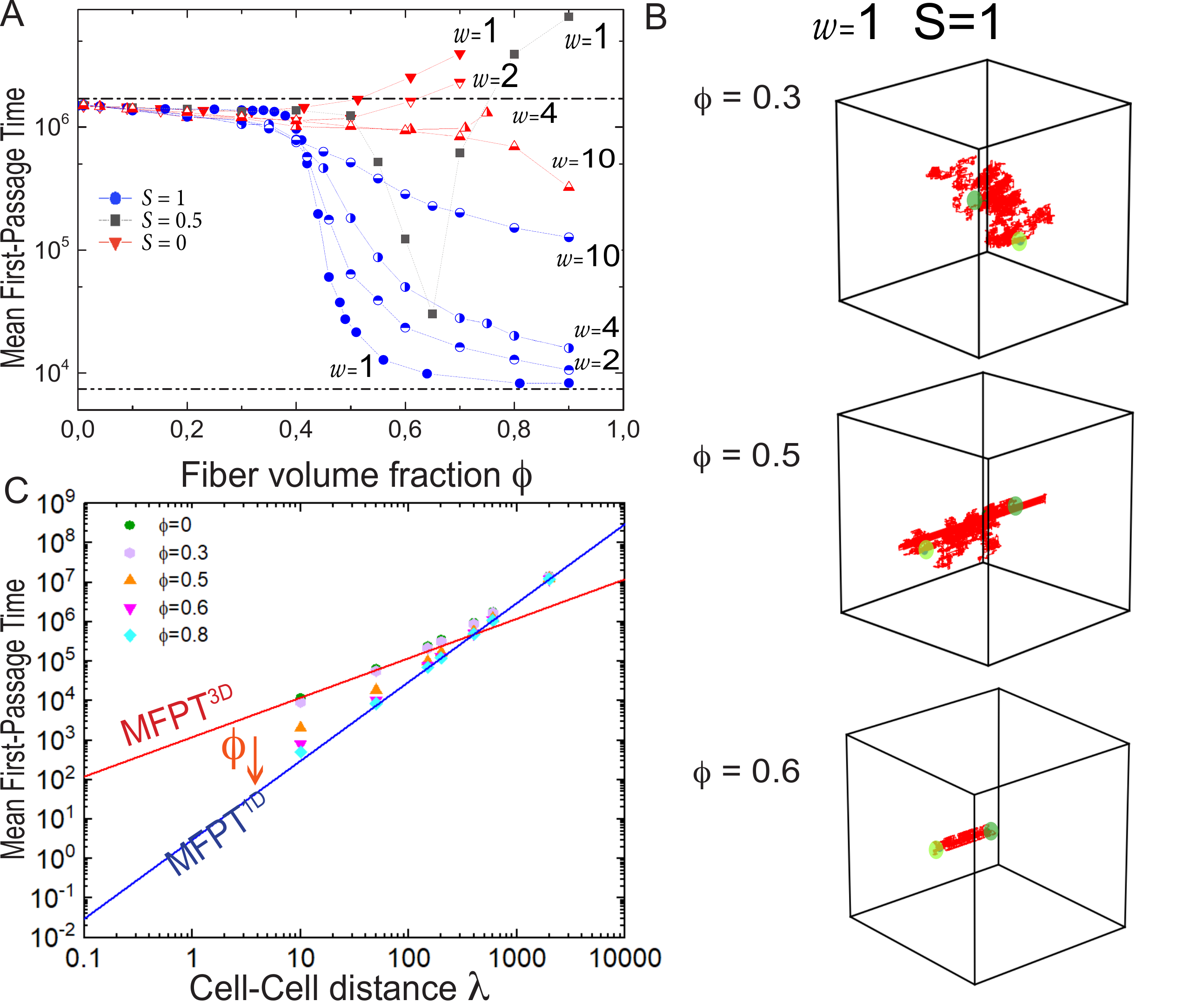} 
\caption{Molecular dynamics of target finding. A) MFPT of target finding as a function of the fiber volume fraction for different nematic order parameters ($S=0,0.5,1$) and fiber thicknesses ($w$). For $S=1$ the finding time exhibits a decay for all fiber thicknesses $w$. When $S=0$ the finding time shows a non-monotonic behavior with an initial decrease for intermediate values of $\phi$. As the fiber volume fraction approaches the critical volume fraction, the finding time increases until it eventually diverges at the percolation threshold. The non-monotonic behavior becomes stronger for the intermediate order parameter value $S=0.5$. Upper horizontal line accounts for MFPT$^{3D}=1.5 \cdot 10^{6}$ and the lower horizontal line is MFPT$^{1D}=7500$. B) Trajectories in the aligned system showing different dynamics at fiber volume fraction below and above the critical fiber fraction $\phi_C^{2D}$. Dark and light green dots denote the initial and final position for simulations $4\cdot 10^{4}$  time steps long, respectively. C) The MFPT as a function of the cell-target distance $\lambda$ for $S=1$. For several chosen cell-cell distances we run simulations with $\phi=0, 0.3, 0.5, 0.6,$ and $0.8$ ({\color{black}colored symbols}) . Simulation parameters: $L_x=L_y=L_z=100$ and $\lambda=50$ for (A) and (B). $L_x=L_z=20$ and $L_y=2\lambda$ for (C).}
\end{figure}
\end{center}



\begin{center}
\begin{figure}
\centering
\includegraphics[width=17.5cm, height=6cm]{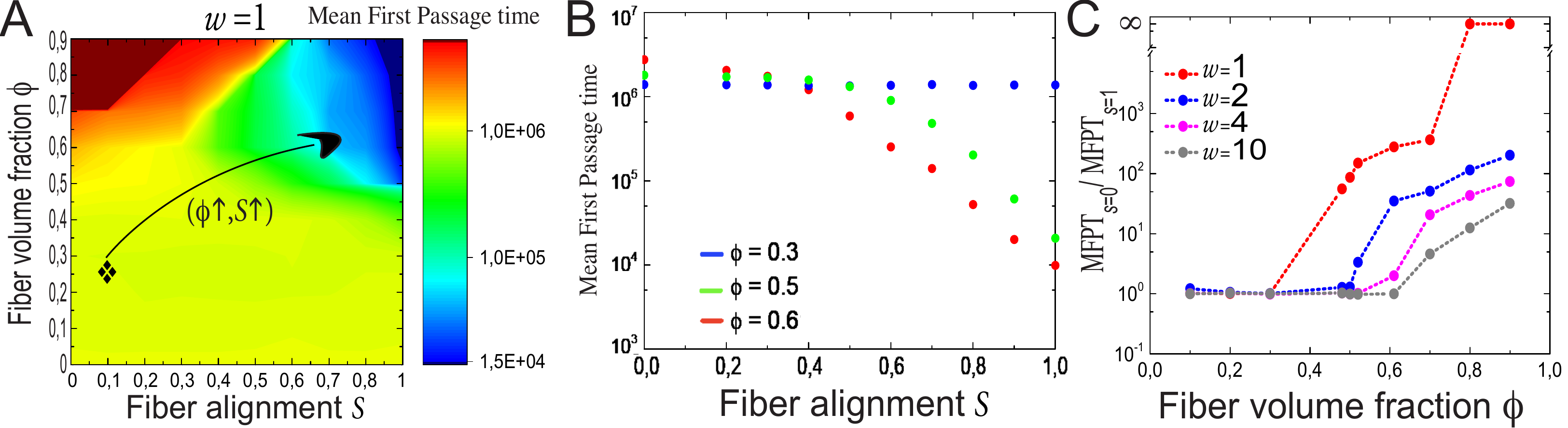} 
\caption{Effects of fiber isotropy on the MFPT of target finding. A) Color map of the MFPT as a function of the fiber volume fraction $\phi$ and nematic order parameter $S$. Dark blue depicts the MFPT that effectively follows a 1D dynamics, whereas dark red represents the region with divergent finding times, occurring for 3D search dynamics. B) MFPT as a function of the system's isotropy for three fiber volume fractions. C) The efficiency of fiber alignment given by the ratio between the MFPT in the isotropic ($S=0$) and aligned ($S=1$) configuration is shown as a function of the fiber volume fraction and for different fiber thicknesses. The ratio increases with the fiber volume fraction. Simulation parameters: $L_x=L_y=L_z=100$.}
\end{figure}
\end{center}

\begin{center}
\begin{figure}
\centering
\includegraphics[width=10cm, height=13cm]{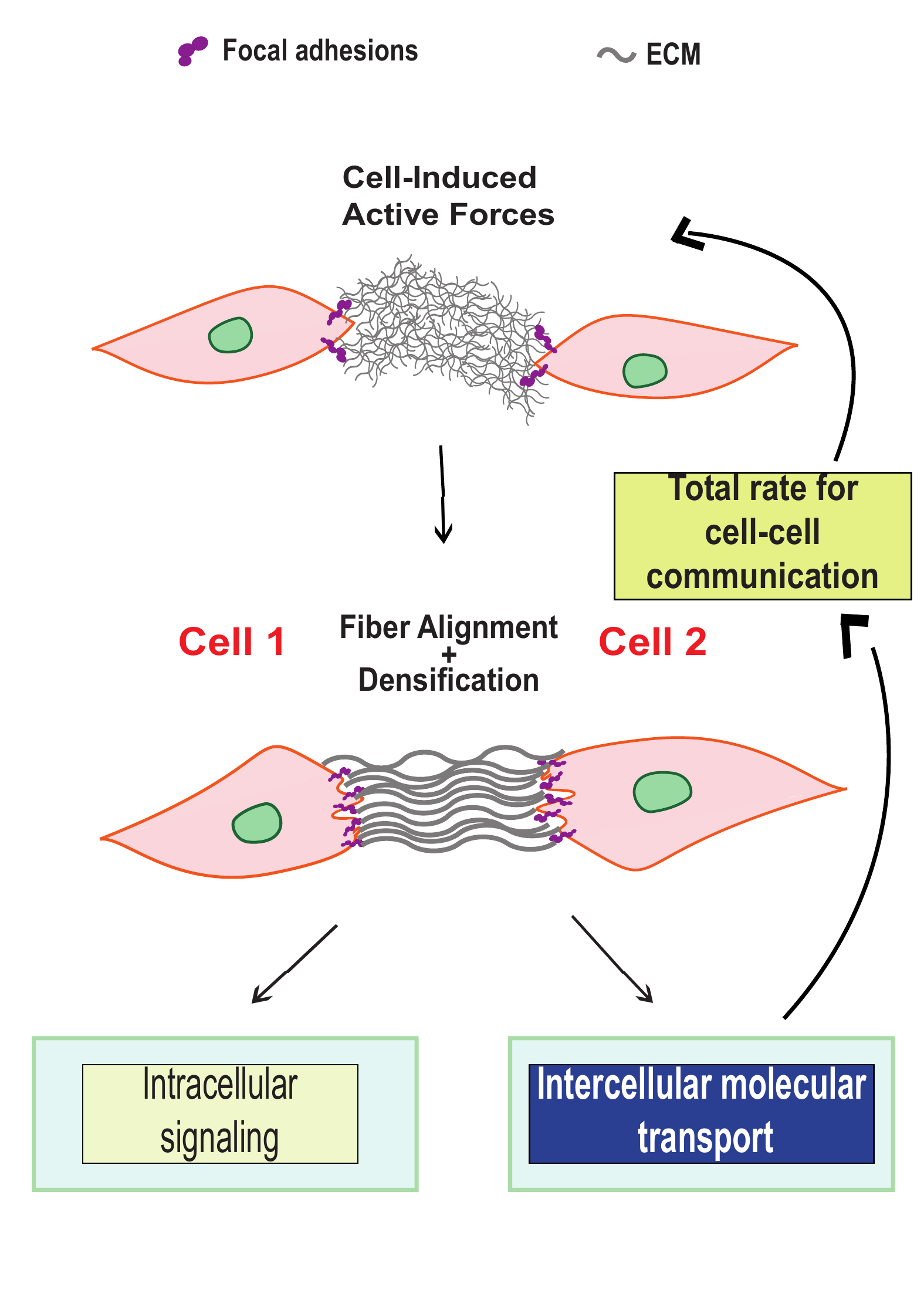} 
\caption{Schematic flow of the signaling transduction mechanism that links biochemical with mechanosensing pathways. Cells 1 and 2 induce mechanical forces on the ECM leading to densification and alignment of the fibers elements. As indicated by our simulations, both densification and fiber alignment support faster transport of molecules by changing molecules dynamics from 3D diffusion to a more efficient 1D process. The total time for cell-cell communication constitutes the time molecules need to be transported from cell to cell (right green box), plus the time required to synthesize molecules, secrete them out of the cell, and the time to activate the signaling pathways (left green box). }
\end{figure}
\end{center}

\end{document}